\documentclass[preprintnumbers, prd, twocolumn, showpacs, floatfix,
preprintnumbers, letterpaper, superscriptaddress,nofootinbib]{revtex4}
\usepackage{amsfonts}
\usepackage{eurosym}
\usepackage[dvips]{graphicx}
\usepackage{epsf}
\usepackage{amsmath}
\usepackage{amssymb}
\usepackage{graphicx}
\usepackage{dcolumn}
\usepackage{bm}
\usepackage{color,wasysym}

\newcommand {\lla} {\ {\raise-.5ex\hbox{$\buildrel<\over\sim$}}\
}

\def\be{\begin{equation}}
\def\ee{\end{equation}}
\def\bea{\begin{eqnarray}}
\def\eea{\end{eqnarray}}

\def\eqi{\begin{equation}}
\def\eqf{\end{equation}}
\def\eqia{\begin{eqnarray}}
\def\eqfa{\end{eqnarray}}

\def\ee{e^2}

\begin{document}

\title{New Schwarzschild-like solutions in $f(T)$ gravity through Noether
symmetries}
\author{A. Paliathanasis}
\email{anpaliat@phys.uoa.gr }
\affiliation{Faculty of Physics, Department of Astrophysics - Astronomy - Mechanics
University of Athens, Panepistemiopolis, Athens 157 83, Greece}
\affiliation{Dipartimento di Fisica, Universita' di Napoli, ``Federico II'', Compl. Univ.
di Monte S. Angelo, Edificio G, Via Cinthia, I-80126, Napoli, Italy}
\affiliation{INFN Sez. di Napoli, Compl. Univ. di Monte S. Angelo, Edificio G, Via
Cinthia, I-80126, Napoli, Italy}
\author{S. Basilakos}
\email{svasil@Academyofathens.gr}
\affiliation{Academy of Athens, Research Center for Astronomy and Applied Mathematics,
Soranou Efesiou 4, 11527, Athens, Greece}
\author{E. N. Saridakis}
\email{Emmanuel\_Saridakis@baylor.edu}
\affiliation{Physics Division, National Technical University of Athens, 15780 Zografou
Campus, Athens, Greece}
\affiliation{Instituto de F\'{\i}sica, Pontificia Universidad de Cat\'olica de Valpara%
\'{\i}so, Casilla 4950, Valpara\'{\i}so, Chile}
\author{S. Capozziello}
\email{capozzie@na.infn.it}
\affiliation{Dipartimento di Fisica, Universita' di Napoli, ``Federico II'', Compl. Univ.
di Monte S. Angelo, Edificio G, Via Cinthia, I-80126, Napoli, Italy}
\affiliation{INFN Sez. di Napoli, Compl. Univ. di Monte S. Angelo, Edificio G, Via
Cinthia, I-80126, Napoli, Italy}
\affiliation{Gran Sasso Science Institute (INFN), Viale F. Crispi, 7, I-67100, L'Aquila,
Italy}
\author{K. Atazadeh}
\email{atazadeh@azaruniv.ac.ir}
\affiliation{Department of Physics, Azarbaijan Shahid Madani University , Tabriz,
53714-161 Iran}
\affiliation{Research Institute for Astronomy and Astrophysics of Maragha (RIAAM),
Maragha 55134-441, Iran}
\author{F. Darabi}
\email{f.darabi@azaruniv.edu}
\affiliation{Department of Physics, Azarbaijan Shahid Madani University , Tabriz,
53714-161 Iran}
\author{M. Tsamparlis}
\email{mtsampa@phys.uoa.gr}
\affiliation{Faculty of Physics, Department of Astrophysics - Astronomy - Mechanics
University of Athens, Panepistemiopolis, Athens 157 83, Greece}

\begin{abstract}
Spherically symmetric solutions for $f(T)$ gravity models are derived by the
so called Noether Symmetry Approach. First, we present a full set of Noether
symmetries for some minisuperspace models. Then, we compute analytical
solutions and find that spherically symmetric solutions in $f(T)$ gravity
can be recast in terms of Schwarzschild-like solutions modified by a
distortion function depending on a characteristic radius. The obtained
solutions are more general than those obtained by the usual solution methods.
\end{abstract}

\maketitle



\section{Introduction}

Modified gravity (see for instance \cite{Capozziello:2011et}) and dark
energy model (see for instance \cite{Copeland:2006wr}) are known as the two
basic approaches to describe the observed acceleration of the universe. The
former dealt with the modification of Einstein's General Relativity itself
whereas the latter suggests some modifications of the cosmic fluid in
Einstein's General Relativity. Essentially, the two approaches consider
modifications in the l.h.s (the former) or in r.h.s. (the latter) of the
cosmological field equations with respect to the picture of the Cosmological
Standard Model. Amongst the variety of modified gravity theories, $f(T)$
gravity has recently received considerable amount of interest and attention.
It is based on the old formulation of ``Teleparallel Equivalent of General
Relativity'' (TEGR) \cite{ein28,Hayashi79,Maluf:1994ji} which instead of the
torsion-less Levi-Civita connection uses the curvature-less Weitzenb{\"{o}}%
ck one, however instead of the torsion scalar $T$ it uses $f(T)$ extensions
in the Lagrangian where $f$ is a function of $T$ \cite%
{Ferraro:2006jd,Ben09,Linder:2010py}.

Although TEGR coincides completely with General Relativity both at the
background and perturbation levels, $f(T)$ gravity proves to exhibit novel
structural and phenomenological features. In particular, imposing a
cosmological background, one can extract various cosmological solutions,
consistent with the observable behavior \cite%
{Ferraro:2006jd,Ben09,Linder:2010py,Myrzakulov:2010vz,Ferraro001,
Wu:2010mn,Iorio:2012cm}. Furthermore, imposing spherical geometry one can
investigate the spherical, black-hole solutions for $f(T)$ gravity \cite%
{Wang:2011xf,Boehmer:2011gw,Daouda001, Gonzalez:2011dr}. These features mean
that $f(T)$ gravity can be interesting, in principle, both at cosmological
and at astrophysical levels.

A crucial point in the context of $f(T)$ gravity is about the allowed
classes of $f(T)$ models. The aforementioned cosmological and spherical
solutions lead to several viable models, although cosmological observations
\cite{Wu:2010mn,BasNes13} as well as Solar System tests \cite{Iorio:2012cm}
indicate that $f(T)$ must be close to the linear form. Thus, in \cite%
{Wei:2011aa,Atazadeh:2011aa,BB1} the authors followed the Noether Symmtery
Approach \cite{Capozziello:1996bi} in order to constrain the allowed $f(T)$
forms that are compatible with the Lema\^\i tre-Robertson-Walker (FLRW)
geometry. Such an approach which is used to fully solve dynamics and also to
determine exactly the corresponding Lagrangian, allowing for the Noether
currents in a given geometry, is very powerful and can be applied in every
gravitational or field-theoretical scenario \cite%
{Capozziello:1996bi,BB1,Capozziello:1999xs,
Szydlowski:2005nu,Camci:2007zz,Capozziello:2008ch,
Vakili:2008ea,Capozziello:2009te,Zhang:2009mm,Basilakos:2011rx,BB,
Kucukakca:2013mya}.

The Noether Symmetry Approach has a deep physical content, since it offers a
theoretical justification for the specific Lagrangian form, instead of
fixing it by hand or by observations. Additionally, in modified
gravitational theories, where the Birkhoff theorem is not guaranteed, the
Noether approach can lead to new solution subclasses that cannot be obtained
by the vacuum field equations \cite{Capozziello:2012iea}.

In the present work, we apply this technique with the aim to derive new
spherically-symmetric solutions for $f(T)$ gravity. Specifically, in Section %
\ref{model}, we briefly present $f(T)$ gravity, while in Section \ref%
{Lagrformul} we construct the corresponding generalized Lagrangian
formulation. In Section \ref{Noethercond}, we analyze the main properties of
the Noether Symmetry Approach for $f(T)$ gravity, general geometry, and the
particular case of spherically symmetric geometry. Then, in Section \ref%
{Analysol}, we use these results in order to obtain all possible solutions,
including novel ones that could not be obtained by the standard approach.
Finally, we exhibit our conclusions in Section \ref{conclusions}.

\section{$f(T)$ gravity}

\label{model}

Let us review now the basic assumptions of $f(T)$ gravity. The notation is
as follows: Greek indices $\mu, \nu,$... and capital Latin indices $A, B, $%
... run over all coordinate and tangent space-time 0, 1, 2, 3, while lower
case Latin indices (from the middle of the alphabet) $i, j,...$ and lower
case Latin indices (from the beginning of the alphabet) $a,b, $... run over
spatial and tangent space coordinates 1, 2, 3, respectively.

In the theory of ``teleparallel'' gravity, as well as in its $f(T)$
extension, the dynamical variable is the vierbein field ${\mathbf{e}_A(x^\mu)%
}$. This forms an orthonormal basis for the tangent space at each point $%
x^\mu$ of the manifold, that is $\mathbf{e} _A\cdot\mathbf{e}_B=\eta_{AB}$,
where $\eta_{AB}=\mathrm{diag} (-1,+1,+1,+1)$. Additionally, the vector $%
\mathbf{e}_A$ can be analyzed with the use of its components $e_A^\mu$ in a
coordinate basis, namely $\mathbf{e}_A=e^\mu_A\partial_\mu $. Finally, in
such a construction, the metric tensor is obtained from the dual vierbein as
\begin{equation}  \label{metrdef}
g_{\mu\nu}(x)=\eta_{AB}\, e^A_\mu (x)\, e^B_\nu (x).
\end{equation}
Contrary to General Relativity, which uses the torsion-less Levi-Civita
connection, in the present gravitational formulation one uses the
curvature-less Weitzenb\"{o}ck connection $\overset{\mathbf{w}}{\Gamma}%
^\lambda_{\nu\mu}\equiv e^\lambda_A\: \partial_\mu e^A_\nu$ \cite{Weitzenb23}%
, and defines the torsion tensor as
\begin{equation}  \label{torsion2}
{T}^\lambda_{\:\mu\nu}=\overset{\mathbf{w}}{\Gamma}^\lambda_{ \nu\mu}-%
\overset{\mathbf{w}}{\Gamma}^\lambda_{\mu\nu} =e^\lambda_A\:(\partial_\mu
e^A_\nu-\partial_\nu e^A_\mu).
\end{equation}
Furthermore, the contorsion tensor
is defined as
\begin{equation}
K^{\mu\nu}_{\:\:\:\:\rho}\equiv-\frac{1}{2}\Big(T^{\mu\nu}_{ \:\:\:\:\rho}
-T^{\nu\mu}_{\:\:\:\:\rho}-T_{\rho}^{\:\:\:\:\mu\nu}\Big),
\end{equation}
and for convenience, we also introduce the tensor
\begin{equation}
S_\rho^{\:\:\:\mu\nu}\equiv\frac{1}{2}\Big(K^{\mu\nu}_{\:\:\:\:\rho}
+\delta^\mu_\rho \:T^{\alpha\nu}_{\:\:\:\:\alpha}-\delta^\nu_\rho\:
T^{\alpha\mu}_{\:\:\:\:\alpha}\Big).
\end{equation}
Using these quantities one can define the teleparallel Lagrangian, which is
the torsion scalar \cite{Hayashi79,Maluf:1994ji}, as\footnote{%
A discussion concerning the role of Torsion in General Relativity can be
found in Basilakos et al. \cite{BB1}.}
\begin{equation}  \label{telelag}
T\equiv S_\rho^{\:\:\:\mu\nu}\:T^\rho_{\:\:\:\mu\nu}=\frac{1}{4} T^{\rho \mu
\nu} T_{\rho \mu \nu} +\frac{1}{2}T^{\rho \mu \nu }T_{\nu \mu\rho } -T_{\rho
\mu }^{\ \ \rho }T_{\ \ \ \nu }^{\nu \mu }.
\end{equation}
In summary, in the present formalism, all the information concerning the
gravitational field is included in the torsion tensor ${T}%
^\lambda_{\:\mu\nu} $, and the torsion scalar $T$ arises from it in a
similar way as the curvature scalar arises from the curvature Riemann tensor
in General Relativity.

While in the teleparallel equivalent of General Relativity (TEGR) the action
is just $T$, the idea of $f(T)$ gravity is to generalize $T$ to a function $%
f(T)$. This is similar in spirit to the generalization of the Ricci scalar $%
R $ in the Einstein-Hilbert action of General Relativity, to a function $f(R)
$ \cite{Capozziello:2011et}. In particular, the action of $f(T)$ gravity is
written as
\begin{eqnarray}  \label{action}
I = \frac{1}{16\pi G}\int d^4x e \left[f(T)\right],
\end{eqnarray}
where $e = \text{det}(e_{\mu}^A) = \sqrt{-g}$, $G$ is the Newton's constant,
and we have set the light speed to 1. We remark here that in some works in
the literature, $T$ is generalized to $T+f(T)$, however in the present
analysis it is proved more convenient to use the above ansatz. Therefore,
TEGR (and thus General Relativity) is restored when $f(T)=T$ (plus a
constant if we consider also the cosmological constant term).

Variation of the action (\ref{action}) with respect to the vierbein gives
the equations of motion
\begin{eqnarray}  \label{eom}
&&e^{-1}\partial_{\mu}(ee_A^{\rho}S_{\rho}{}^{\mu\nu})f_{T}
-e_{A}^{\lambda}T^{\rho}{}_{\mu\lambda}S_{\rho}{}^{\nu\mu}f_{T}  \notag \\
&&\ \ \, + e_A^{\rho}S_{\rho}{}^{\mu\nu}\partial_{\mu}({T})f_{TT}+\frac{1}{4}
e_ { A } ^ { \nu }f({T}) = 4\pi Ge_{A}^{\rho}\overset{\mathbf{em}}{T}%
_{\rho}{}^{\nu},
\end{eqnarray}
where $f_{T}$ and $f_{TT}$ denote the first and second derivatives of the
function $f(T)$ with respect to $T$, respectively. Finally, the tensor $%
\overset{\mathbf{em}}{T}_{\rho}{}^{\nu}$ stands for the usual
energy-momentum tensor of perfect fluid matter.

\section{Generalized Lagrangian formulation of $f(T)$ gravity}

\label{Lagrformul}

In this section, following the technique described in \cite{BB,TsamGE}, we
provide a generalized Lagrangian formulation in order to construct a theory
of $f(T)$ gravity. Specifically, the gravitational field is driven by the
Lagrangian density $f(T)$ in (\ref{action}), which can be generalized
through the use of a Lagrange multiplier. In particular, we can write it as
\begin{equation}
L\left( x^{k},x^{\prime k},T\right) =2f_{T}\bar{\gamma}_{ij}\left(
x^{k}\right) x^{\prime i}x^{\prime j}+M\left( x^{k}\right) \left(
f-Tf_{T}\right) ,  \label{L1}
\end{equation}%
where $x^{\prime }=\frac{dx}{d\tau }$, $M(x^{k})$ is the Lagrange multiplier
and $\bar{\gamma}_{ij}$ is a second rank tensor which is related to the
frame [one can use $eT(x^{k},x^{\prime k})$] of the background spacetime. In
the same lines, the Hamiltonian of the system is written as
\begin{equation}
H\left( x^{k},x^{\prime k},T\right) =2f_{T}\bar{\gamma}_{ij}\left(
x^{k}\right) x^{\prime i}x^{\prime j}-M\left( x^{k}\right) \left(
f-Tf_{T}\right) \;.  \label{L2}
\end{equation}

In this case, the system is autonomous and because of that $\partial _{\tau
} $ is a Noether symmetry with corresponding Noether integral the
Hamiltonian $H$. Additionally, since the coupling function $M$ is a function
of $x^{k}$, it is implied that the Hamiltonian (\ref{L2}) vanishes \cite%
{Paliathanasis2013GRG}.

In this framework, considering $\{x^{k},T\}$ as the canonical variables of
the configuration space, we can derive, after some algebra, the general
field equations of $f(T)$ gravity. Indeed, starting from the Lagrangian (\ref%
{L1}), the Euler-Lagrange equations
\begin{equation}
\frac{\partial L}{\partial T}=0,\;\;\;\;\frac{d}{d\tau }\left( \frac{%
\partial L}{\partial x^{\prime k}}\right) -\frac{\partial L}{\partial x^{k}}%
=0\,,  \label{Lf.01}
\end{equation}%
give rise to
\begin{equation}
f_{TT}\left( 2\bar{\gamma}_{ij}x^{\prime i}x^{\prime j}-MT\right) =0,
\label{Lf.04}
\end{equation}%
\begin{equation}
x^{i\prime \prime }+{\bar \Gamma}_{jk}^{i}x^{j\prime }x^{k\prime }+\frac{%
f_{TT}}{f_{T}}x^{i\prime }T^{\prime }-M^{,i}\frac{\left( f-Tf_{T}\right) }{%
4f_{T}}=0\;.  \label{Lf.06}
\end{equation}%
We mention here that, for convenience, the functions ${\bar \Gamma}_{jk}^{i}$
are considered: they are exactly the Christoffel symbols for the metric $%
\bar{\gamma}_{ij}$. Therefore, the system is determined by the two
independent differential equations (\ref{Lf.04}),(\ref{Lf.06}), and the
Hamiltonian constrain $H=0$ where $H$ is given by Eq.(\ref{L2}).

The point-like Lagrangian (\ref{L1}) determines completely the related
dynamical system in the minisuperspace $\{x^{k},T \}$, implying that one can
easily recover some well known cases of cosmological interest. In brief,
these are:

\begin{itemize}
\item The static spherically symmetric spacetime:
\begin{eqnarray}  \label{SS}
&&ds^{2}=-a^{2}\left( \tau \right) dt^{2}+\frac{1}{N^{2}\left( a\left( \tau
\right) ,b\left( \tau \right) \right) }d\tau ^{2}  \notag \\
&&\ \ \ \ \ \ \ \ \, +b^{2}\left( \tau \right) \left( d\theta ^{2}+\sin
^{2}\theta d\phi ^{2}\right),
\end{eqnarray}
arising from the diagonal vierbein \footnote{%
Note that, in general, one can choose a non-diagonal vierbein, giving rise
to the same metric through (\ref{metrdef}). However for the sake of
simplicity, we remain in the diagonal case which is capable of revealing the
main features of the solutions \cite{Wang:2011xf,Gonzalez:2011dr,Wei:2011aa}.%
}
\begin{equation}  \label{SSA}
e_{i}^{A}=\left( a\left( \tau \right) ,\frac{1}{N\left( a\left( \tau \right)
,b\left( \tau \right) \right) },b\left( \tau \right) ,b\left( \tau \right)
\sin \theta \right) \;,
\end{equation}
where $a(\tau)$ and $b(\tau)$ are functions which need to be determined.
Therefore, the line element of $\bar{\gamma}_{ij}$ and $M\left( x^{k}\right)$
are given by
\begin{eqnarray}  \label{SSA1}
&&ds_{\bar{\gamma}}^{2}=N\left( 2b~da~db+a~db^{2}\right),  \notag \\
&&M(a,b)=\frac{ab^{2}}{N}.
\end{eqnarray}

\item The flat Friedmann-Lema\^\i tre-Robertson-Walker spacetime with
Cartesian coordinates:
\begin{equation}
ds^{2}=-dt^{2}+a^{2}\left( t\right) \left( dx^{2}+dy^{2}+dz^{2}\right),
\end{equation}
arising from the vierbein
\begin{equation}
e_{i}^{A}=\left( 1,a\left( t\right) ,a\left( t\right) ,a\left( t\right)
\right) \;,
\end{equation}
where $t$ is the cosmic time and $a(t)$ is the scale factor of the universe.
In this case we have
\begin{eqnarray}
&&ds_{\bar{\gamma}}^{2}=3a~da^{2},  \notag \\
&&M(a)=a^{3}(t).
\end{eqnarray}

\item The Bianchi type I spacetime:
\begin{eqnarray}
&&\!ds^{2}=-\frac{1}{N^{2}\left( a\left( t\right) ,\beta \left( t\right)
\right) }dt^{2}  \notag \\
&&\ \ \ \ \ \, \ \ \, +a^{2}\left( t\right) \left[ e^{-2\beta \left(
t\right) }dx^{2}+e^{\beta \left( t\right) }\left( dy^{2}+dz^{2}\right) %
\right],\
\end{eqnarray}%
arising from the vierbein
\begin{equation}
e_{i}^{A}=\left( \frac{1}{N\left( a\left( t\right) ,\beta \left( t\right)
\right) },a(t)e^{-\beta (t)},a(t)^{\frac{\beta (t)}{2}},a(t)^{\frac{\beta (t)%
}{2}}\right) \;.
\end{equation}%
In this case, we obtain
\begin{eqnarray}
&&ds_{\bar{\gamma}}^{2}=N\left( -4ada^{2}+a^{3}d\beta ^{2}\right) ,  \notag
\\
&&M(a,\beta )=\frac{a^{3}(t)}{N}.
\end{eqnarray}
\end{itemize}

In the present work we will focus on the static spherically-symmetric metric
deriving new spherically symmetric solutions for $f(T)$ gravity. In
particular, we look for Noether symmetries in order to reveal the existence
of analytical solutions.

\section{The Noether Symmetry Approach for $f(T)$ gravity}

\label{Noethercond}

The aim is now to extend results in \cite{Wei:2011aa} and \cite{BB1} by
applying the Noether Symmetry Approach \cite{TsamGE} to a general class of $%
f(T)$ gravity models where the corresponding Lagrangian of the field
equations is given by Eq.(\ref{L1}). First of all, we perform the analysis
for arbitrary spacetimes, and then we focus on static spherically-symmetric
geometries.

\subsection{Searching for Noether point symmetries in general spacetimes}

The Noether symmetry condition for the Lagrangian (\ref{L1}) is given by
\begin{equation}
X^{\left[ 1\right] }L+L\xi ^{\prime }=g^{\prime },  \label{LL2}
\end{equation}%
where the generator $X^{\left[ 1\right] }$ is written as
\begin{eqnarray}
&&X^{\left[ 1\right] }=\xi \left( \tau ,x^{k},T\right) \partial _{\tau
}+\eta ^{k}\left( \tau ,x^{k},T\right) \partial _{i}  \notag \\
&&\ \ \ \ \ \ \ \ \ +\mu \left( \tau ,x^{k},T\right) \partial _{T}+\left(
\eta ^{\prime i}-\xi ^{\prime }x^{\prime i}\right) \partial _{x^{\prime i}}.
\end{eqnarray}%
For each term of the Noether condition (\ref{LL2}) for the Lagrangian (\ref%
{L1}) we obtain
\begin{eqnarray*}
X^{\left[ 1\right] }L &=&2f_{T}\bar{g}_{ij,k}\eta ^{k}x^{\prime i}x^{\prime
j}+M_{,k}\eta ^{k}\left( f-Tf_{T}\right) \\
&&+2f_{TT}\mu \bar{g}_{ij}x^{\prime i}x^{\prime j}-Mf_{TT}\mu \\
&&+4f_{T}\bar{g}_{ij}x^{\prime i}\left( \eta _{,\tau }^{j}+\eta
_{,k}^{j}x^{\prime k}+\eta _{,T}^{j}T^{\prime }\right. \\
&&\left. -\xi _{,\tau }x^{\prime j}-\xi _{,k}x^{\prime j}x^{\prime k}-\xi
_{,T}x^{\prime j}T^{\prime }\right) ,
\end{eqnarray*}%
\begin{eqnarray*}
&&L\xi ^{\prime }=\left[ 2f_{T}\bar{g}_{ij}x^{\prime i}x^{\prime j}+M\left(
x^{i}\right) \left( f-Tf_{T}\right) \right] \\
&&\ \ \ \ \ \ \ \ \ \cdot \left( \xi _{,\tau }+\xi _{,k}x^{\prime k}+\xi
_{,T}T^{\prime }\right) ,
\end{eqnarray*}%
\begin{equation*}
g^{\prime }=g_{,\tau }+g_{,k}x^{\prime k}+g_{,T}T^{\prime }\;.
\end{equation*}%
Inserting these expressions into (\ref{LL2}) we find the Noether symmetry
conditions
\begin{equation}
\xi _{,k}=0~,~\xi _{,T}=0~,~g_{,T}=0~,~\eta _{,T}=0,
\end{equation}%
\begin{equation}
4f_{T}\bar{\gamma}_{ij}\eta _{,\tau }^{k}=g_{,k},  \label{LL3}
\end{equation}%
{\small {\
\begin{equation}
M_{,k}\eta ^{k}\left( f-Tf_{T}\right) -MTf_{TT}\mu +\xi _{,\tau }M\left(
f-Tf_{T}\right) -g_{,\tau }=0,  \label{LL4}
\end{equation}%
}}
\begin{equation}
2f_{T}\bar{\gamma}_{ij,k}\eta ^{k}+2f_{TT}\mu \bar{\gamma}_{ij}+4f_{T}\bar{%
\gamma}_{ij}\eta _{,k}^{j}-2f_{T}\bar{\gamma}_{ij}\xi _{,\tau }=0\;.
\label{LL5}
\end{equation}%
Notice that conditions $\eta _{,T}=g_{,T}=0$ imply, through Eq.(\ref{LL3}),
that $\eta _{,\tau }^{k}=g_{,k}=0$. Also, Eq.(\ref{LL5}) takes the form
\begin{equation}
L_{\eta }\bar{\gamma}_{ij}=\left( \xi _{,\tau }-\frac{f_{TT}}{f_{T}}\mu
\right) \bar{\gamma}_{ij},  \label{LL6}
\end{equation}%
where $L_{\eta }\bar{\gamma}_{ij}$ is the Lie derivative with respect to the
vector field $\eta ^{i}(x^{k})$. Furthermore, from (\ref{LL6}) we deduce
that $\eta ^{i}$ is a Conformal Killing Vector of the metric $\bar{\gamma}%
_{ij}$, and the corresponding conformal factor is
\begin{equation}
2\bar{\psi}\left( x^{k}\right) =\xi _{,\tau }-\frac{f_{TT}}{f_{T}}\mu =\xi
_{,\tau }-S(\tau ,x^{k})\;.  \label{LL7}
\end{equation}%
Finally, utilizing simultaneously Eqs.(\ref{LL4}), (\ref{LL6}), (\ref{LL7})
and the condition $g_{,\tau }=0$, we rewrite (\ref{LL4}) as
\begin{equation}
M_{,k}\eta ^{k}+\left[ 2\bar{\psi}+\left( 1-\frac{Tf_{T}}{f-Tf_{T}}\right) S%
\right] M=0\;.  \label{LL9}
\end{equation}%
Considering that $S=S(x^{k})$ and using the condition $g_{,\tau }=0$, we
acquire $\xi _{,\tau }=2\bar{\psi}_{0},\bar{\psi}_{0}\in \mathbb{R}$ with $%
S=2(\bar{\psi}_{0}-\bar{\psi})$. At this point, we have to deal with the
following two situations:

Case 1. In the case of $S=0$, the symmetry conditions are
\begin{eqnarray}
&&L_{\eta }\bar{\gamma}_{ij}=2\bar{\psi}_{0}\bar{\gamma}_{ij},  \notag \\
&&M_{,k}\eta ^{k}+2\bar{\psi}_{0}M=0,
\end{eqnarray}%
implying that the vector $\eta ^{i}(x^{k})$ is a Homothetic Vector of the
metric $~\bar{\gamma}_{ij}$. The latter means that for arbitrary $f\left(
T\right) \neq T^{n}$ functional forms, our dynamical system could possibly
admit extra (time independent) Noether symmetries.

Case 2. If $S\neq 0$ then Eq. (\ref{LL9}) immediately leads to the following
differential equation
\begin{equation}
\frac{Tf_{T}}{ f-Tf_{T}}=C,
\end{equation}
which has the solution
\begin{equation}
f(T)=T^{n}, \;\;\;\;C\equiv \frac{n}{1-n} \;.
\end{equation}
In this context, $\eta ^{i}(x^{k})$ is a Conformal Killing Vector of $\bar{%
\gamma}_{ij}$, and the symmetry conditions become
\begin{eqnarray}
&&L_{\eta }\bar{\gamma}_{ij}=2\bar{\psi}\bar{\gamma}_{ij},  \notag \\
&& M_{,k}\eta ^{k}+\left[ 2\bar{\psi}+\left( 1-C\right) S\right] =0,
\end{eqnarray}
with $S=2(\bar{\psi}_{0}-\bar{\psi})$.

Collecting the above results we can formulate the following proposition:

\textbf{\emph{Lemma:}}

\textit{The general autonomous Lagrangian
\begin{equation*}
L\left( x^{k},x^{\prime k},T\right) =2f_{T}\bar{\gamma}_{ij}\left(
x^{k}\right) x^{\prime i}x^{\prime j}+M\left( x^{k}\right) \left(
f-Tf_{T}\right)
\end{equation*}%
admits extra Noether symmetries as follows: }

\begin{enumerate}
\item \textit{If $f(T)$ is an arbitrary function of $T$, then the symmetry
vector is written as
\begin{equation*}
X^{[1]}=\left( 2\psi _{0}\tau +c_{1}\right) \partial _{\tau }+\eta
^{i}\left( x^{k}\right) \partial _{i},
\end{equation*}%
where $\eta ^{i}\left( x^{k}\right) $ is a Homothetic Vector of the metric $%
\bar{\gamma}_{ij}$ and the following condition holds
\begin{equation*}
M_{,k}\eta ^{k}+2\bar{\psi}_{0}M=0\;.
\end{equation*}%
Note that if $\eta ^{i}$ is a Killing Vector (or Homothetic Vector) then $%
\psi _{0}=0$ (or $\psi _{0}=1$). }

\item \textit{If $f(T)$ is a power law, namely $T^{n}$, then we have the
extra symmetry vector
\begin{equation*}
X^{[1]}=\left( 2\bar{\psi}_{0}\tau \right) \partial _{\tau }+\eta ^{i}\left(
x^{k}\right) \partial _{i}+\frac{\left( 2\bar{\psi}_{0}-2\bar{\psi}\right) }{%
C}T\partial _{T},
\end{equation*}%
where $C=\frac{n}{1-n}$, $\eta ^{i}$ is a Conformal Killing Vector of the
metric $\bar{\gamma}_{ij}$ with conformal factor $\bar{\psi}\left(
x^{k}\right) $ and the following condition holds
\begin{equation*}
M_{,k}\eta ^{k}+\left[ 2\bar{\psi}+\left( 1-C\right) S\right] =0,
\end{equation*}%
with $S=2(\bar{\psi}_{0}-\bar{\psi})$. }

\textit{In both cases the corresponding gauge function is a constant. }
\end{enumerate}

\subsection{ Noether symmetries of the field equations in static spherically
symmetric spacetimes}

\label{spergem}

Let us now apply the results of the general Noether analysis of the previous
subsection, to the specific case of static spherically-symmetric geometry,
which is the subject of interest of the present work. Thus, from now on we
focus on the metric (\ref{SS}), 
that is the vierbein (\ref{SSA}). 

\begin{table}[ht]
\caption{Noether symmetries and integrals for arbitrary $f(T)$.}\tabcolsep %
6pt
\begin{tabular}{ccc}
\hline
$N(a,b)$ & Symmetry & Integral \\ \hline\hline
$\frac{1}{a^{3}}N_{1}\left( a^{2}b\right) $ & $-\frac{a}{2b^{3}}\partial
_{a}+\frac{1}{b^{2}}\partial _{b}$ & $\frac{N_{1}\left( a^{2}b\right) }{%
2a^{3}b^{2}}\left( 2ba^{\prime }+ab^{\prime }\right) f_{T}$ \\
$N_{2}\left( b\sqrt{a}\right) $ & $-2a\partial _{a}+b\partial _{b}$ & $%
N_{2}\left( b\sqrt{a}\right) \left( b^{2}a^{\prime }-abb^{\prime }\right)
f_{T}$ \\
$aN_{3}\left( b\right) $ & $\frac{1}{ab}\partial _{a}~$ & $N_{3}\left(
b\right) b^{\prime }f_{T}$ \\ \hline
\end{tabular}%
\end{table}

\begin{table*}[ht]
\caption{Extra Noether symmetries and integrals for $f(T)=T^{n}$ with $C=%
\frac{n}{1-n}$. The last four lines correspond to the special case where $%
n=1/2$. Notice, that ${\bar \protect\psi}_{5-7}$ are the conformal factors
defined as ${\bar \protect\psi}=\frac{1}{\mathrm{dim}{\bar \protect\gamma}%
_{ij}}\protect\eta^{k}_{;k}$. We notify that the power law case also admits
the Noether symmetries of Table I.}\tabcolsep 6pt
\begin{tabular}{ccc}
\hline
$N(a,b)$ & Symmetry & Integral \\ \hline\hline
arbitrary & $2\bar{\psi}_{0}\tau\partial _{\tau} +\frac{2\bar{\psi}%
_{0}\left( C-1\right) }{2C+1}a\partial _{a}+\frac{2\bar{\psi}_{0}-2\bar{\psi}%
_{4}}{C}T\partial _{T}$ & $2\psi _{0}n\frac{C-1}{1+2C}abN\left( a,b\right)
T^{n-1}b^{\prime }~$ \\
& $-2a\partial _{a}+b\partial _{b}-\frac{2\bar{\psi}_{5}}{C}T\partial _{T}$
& $nN\left( a,b\right) T^{n-1}\left( b^{2}a^{\prime }-abb^{\prime }\right) $
\\
& $-\frac{a}{2}b^{-\frac{3\left( 1+2C\right) }{4C}}\partial _{a}+b^{-\frac{%
3+2C}{4C}}\partial _{b}-\frac{2\bar{\psi}_{6}}{C}T\partial _{T}$ & $\frac{n}{%
2}N\left( a,b\right) T^{n-1}\left( 2b^{\frac{2C-3}{4C}}a^{\prime }+ab^{-%
\frac{3+2C}{4C}}b^{\prime }\right) $ \\
& $a^{-\frac{1}{2C}}b^{-\frac{1+2C}{4C}}\partial _{a}-\frac{2\bar{\psi}_{7}}{%
C}T\partial _{T}~$ & $N\left( a,b\right) na^{-\frac{1}{2C}}b^{-\frac{1+2C}{4C%
}}T^{n-1}b^{\prime }$ \\
&  &  \\
arbitrary & $2\bar{\psi}_{0}\tau\partial _{\tau} +\frac{3\bar{\psi}_{0}}{2}%
a\ln \left( a^{2}b\right) \partial _{a}+\frac{2\bar{\psi}_{0}-2\bar{\psi}%
_{4}^{\prime }}{C}T\partial _{T}$ & $\frac{3}{2}\psi _{0}N\left( a,b\right)
T^{-\frac{1}{2}}ab\ln \left( a^{2}b\right) b^{\prime }$ \\
& $b\partial _{b}-\frac{2\bar{\psi}_{5}}{C}T\partial _{T}$ & $\frac{1}{2}%
N\left( a,b\right) T^{-\frac{1}{2}}\left( b^{2}a^{\prime }+abb^{\prime
}\right) $ \\
& $-a\ln \left( ab\right) \partial _{a}+b\ln b\partial _{b}-\frac{2\bar{\psi}%
_{6}}{C}T\partial _{T}$ & $\frac{1}{2}N\left( a,b\right) T^{-\frac{1}{2}%
}b\left( b\ln b~a^{\prime }-a\ln a~b^{\prime }\right) $ \\
& $a\partial _{a}-\frac{2\bar{\psi}_{7}}{C}T\partial _{T}~$ & $\frac{1}{2}%
N\left( a,b\right) T^{-\frac{1}{2}}ab~b^{\prime }$ \\ \hline
\end{tabular}%
\end{table*}

Armed with the general expressions provided above, we can deduce the Noether
algebra of the metric (\ref{SSA1}). In particular, the Lagrangian (\ref{L1})
and the Hamiltonian (\ref{L2}) become
\begin{equation}
L=2f_{T}N\left( 2ba^{\prime }b^{\prime }+ab^{\prime 2}\right) +M(a,b)\left(
f-f_{T}T\right),
\end{equation}
\begin{equation}
H=2f_{T}N\left( 2ba^{\prime }b^{\prime }+ab^{\prime 2}\right)-M(a,b)\left(
f-f_{T}T\right)\equiv 0 \;,  \label{HLf.06}
\end{equation}
where $M(a,b)$ is given by (\ref{SSA1}). As one can immediately deduce, TEGR
and thus General Relativity is restored as soon as $f(T)=T$, while if $N=1$,
$\tau=r$ and $ab=1$ we fully recover the standard Schwarzschild solution.

Applying the results of the previous subsection in this specific case of
static spherically-symmetric geometry, we determine all the functional forms
of $f(T) $ for which the above dynamical system admits Noether point
symmetries beyond the trivial one $\partial _{\tau}$ related to the energy,
and we summarize the results in Tables I and II. Thus, we can use the
obtained Noether integrals in order to classify the analytical solutions.

\section{New classes of Analytical solutions}

\label{Analysol}

Using the Noether symmetries and the corresponding integral of motions
obtained in the previous section, we can extract all the static
spherically-symmetric solutions of $f(T)$ gravity. We stress that, in this
way, we obtain new solutions, that could not be obtained by the standard
methods applied in \cite{Wang:2011xf,Boehmer:2011gw,Daouda001,
Gonzalez:2011dr}.

Without loss of generality, we choose the conformal factor $N(a,b)$ such as $%
N(a,b)=ab^{2}$ [or equivalently\footnote{%
Since the space is empty, the field equations are conformal invariant,
therefore the results are similar for an arbitrary function $N(a,b)~$ \cite%
{Paliathanasis2013GRG}.} $M(a,b)=1$]. In order to simplify the current
dynamical problem, we consider the coordinate transformation
\begin{equation}
b=\left( 3y\right) ^{\frac{1}{3}},\;\;\;\;\;a=\sqrt{\frac{2x}{\left(
3y\right) ^{\frac{1}{3}}}}\;.  \label{L4S001}
\end{equation}%
Substituting the above variables into the field equations (\ref{Lf.04}), (%
\ref{Lf.06}), (\ref{HLf.06}) we immediately obtain
\begin{eqnarray}
&&x^{\prime \prime }+\frac{f_{TT}}{f_{T}}x^{\prime }T^{\prime }=0,
\label{L4S03} \\
&&y^{\prime \prime }+\frac{f_{TT}}{f_{T}}y^{\prime }T^{\prime }=0,
\label{L4S04} \\
&&H=4f_{T}x^{\prime }y^{\prime }-\left( f-Tf_{T}\right) ,  \label{L4S02}
\end{eqnarray}%
while the torsion scalar (\ref{telelag}) is given by
\begin{equation}
T=4x^{\prime }y^{\prime }\;.  \label{L4S01}
\end{equation}%
Finally, the generalized Lagrangian (\ref{L1}) acquires the simple form
\begin{equation}
L=4f_{T}x^{\prime }y^{\prime }+\left( f-Tf_{T}\right) \;.
\end{equation}

Since the analysis of the previous subsection revealed two classes of
Noether symmetries, namely for arbitrary $f(T)$, and $f(T)=T^{n}$, in the
following subsections we investigate them separately. We would like
to mention that the solutions provided below have been extracted under the
assumption $f_{TT}\neq 0$, that is when $f(T)$ is not a linear function of $T
$. Therefore, our solutions cannot be extrapolated back to the GR solutions
where $f(T)=T$ (additionally note that these two cases exhibit different
phase space and different Noether symmetries, and thus the obtained
solutions do not always have a $f_{TT}\rightarrow0$ limit).

\subsection{Arbitrary $f(T)$}

In the case where $f(T)$ is arbitrary, a ``special solution'' of the system (%
\ref{L4S03})-(\ref{L4S01}) is
\begin{eqnarray}
x\left( \tau \right) &=&c_{1}\tau +c_{2},  \label{L4S5} \\
y\left( \tau \right) &=&c_{3}\tau +c_{4},  \label{L4S6}
\end{eqnarray}%
and the Hamiltonian constraint ($H=0$) reads
\begin{equation}
4c_{1}c_{3}\frac{df}{dT}|_{T=4c_{1}c_{3}}-f+T\frac{df}{dT}%
|_{T=4c_{1}c_{3}}=0,  \label{yLL}
\end{equation}%
where $T=4c_{1}c_{3}$, and $c_{1,..,4}$ are integration constants.  We
mention that the current solution is just one special solution in the case
of arbitrary, non-linear $f(T)$, which is indeed characterized by a constant
$T=4c_{1}c_{3}$. Definitely, the general solution will not have constant $T$.%

Utilizing (\ref{L4S001}), (\ref{L4S5}) and (\ref{L4S6}), we get
\begin{eqnarray}
&&b\left( \tau \right) =3^{\frac{1}{3}}\left( c_{3}\tau +c_{4}\right) ^{%
\frac{1}{3}},  \notag \\
&&a\left( \tau \right) =\frac{\sqrt{6}}{3^{\frac{2}{3}}}\left( c_{1}\tau
+c_{4}\right) ^{\frac{1}{2}}\left( c_{3}\tau +c_{4}\right) ^{\frac{1}{6}}\;.
\end{eqnarray}%
For convenience, we can change variables from $b\left( \tau \right) $ to $r$
according to the transformation $b\left( \tau \right) =r$, where $r$ denotes
the radial variable. Inserting this into the above equations, we conclude
that the spacetime (\ref{SS}) in the coordinates $(t,r,\theta ,\phi )$ can
be written as
\begin{equation}
ds^{2}=-A\left( r\right) dt^{2}+\frac{1}{c_{3}^{2}}\frac{1}{A\left( r\right)
}dr^{2}+r^{2}\left( d\theta ^{2}+\sin ^{2}\theta d\phi ^{2}\right) ,
\label{yLLa}
\end{equation}%
with
\begin{equation}
A\left( r\right) =\frac{2c_{1}}{3c_{3}}r^{2}-\frac{2c_{\mu }}{c_{3}r}%
=\lambda _{A}\left(1-\frac{r_{\star }}{r}\right)R(r),  \label{ALLa}
\end{equation}%
and
\begin{equation}
R(r)=\left( \frac{r}{r_{\star }}\right) ^{2}+\frac{r}{r_{\star }}+1.
\label{RLLa}
\end{equation}%
In these expressions, we have defined $c_{\mu }=c_{1}c_{4}-c_{2}c_{3}$, $%
\lambda _{A}=\left( \frac{8c_{1}c_{\mu }^{2}}{3c_{3}^{3}}\right) ^{1/3}$,
and $r_{\star }=(\frac{3c_{\mu }}{c_{1}})^{1/3}=(\frac{3c_{3}\lambda _{A}}{%
2c_{1}})^{1/2}$ is a characteristic radius with the restriction $c_{\mu
}c_{1}>0.$

As we can observe, if we desire to obtain a Schwarzschild-de Sitter-like metric
we need to select the constant $c_{3}$ such as $c_{3}\equiv 1$. On the other
hand, the function $R(r)$ can be viewed as a distortion factor which
quantifies the deviation from the pure Schwarzschild solution. Thus, the $%
f(T)$ gravity on small spherical scales ($r \rightarrow r_{\star}^{+}$)
tends to create a Schwarzschild solution.

 In order to explore the singularity and horizon features of the
obtained solutions we additionally calculate the Kretschmann scalar [from
the metric (\ref{yLLa}) we calculate the Levi-Civita connection, then the
Riemann tensor $R^{abcd}$, and finally the Kretschmann scalar $\mathcal{K}%
\equiv R^{abcd}R_{abcd}$], obtaining
\begin{eqnarray*}  \label{eq50}
&&\!\!\!\!\!\mathcal{K}=\frac{1}{r^4}\left[c_3^{4}A^{2}_{,rr}r^{4}\right.  \notag \\
&& \left.\ \ \ +4c_3^{4}A^{2}_{,r} r^{2}+4-8c_3^{2}A(r)+4c_3^{4}A(r)^{2}
\right]  \notag \\
&& \, = \frac{4 }{3 r^4}\left\{4 c_3 r \left[c_3 \left(2 c_1^2 r^3-6 c_1
c_{\mu} r^2-3 c_2+6 c_{\mu}^2 r\right)\right.\right.  \notag \\
&&\ \ \ \left.\left.+c_1 (3 c_4-r)\right]+3\right\}.
\end{eqnarray*}
Notice, that in this section we use the general definition 
$\Xi_{,r}=d\Xi/dr$ and $\Xi_{,rr}=d^{2}\Xi^{2}/dr^{2}$.
As we can observe, $\mathcal{K}$ is singular at the origin, and thus the origin
corresponds to a physical singularity, similar to the Schwarzschild
solution. As usual it is hidden behind a horizon at $r = r_{\star}$, in
which the Kretschmann scalar is finite. 

 Let us now examine the remaining physical features of the above
solution. Within this framework, for the observers, $u^{i}u_{i}=-1$, it is
easy to show that the Einstein's tensor becomes
\begin{equation*}
G_{j}^{i}=\mathrm{diag}\left( 2c_{1}c_{3}-\frac{1}{r^{2}},2c_{1}c_{3}-\frac{1%
}{r^{2}},2c_{1}c_{3},2c_{1}c_{3}\right) .
\end{equation*}%
Hence, one can treat the problem at hand using an \textquotedblleft
effective\textquotedblright\ fluid, which can be seen as a dynamical
consequence  of $f(T)$ gravity. Therefore, from the 1+3
decomposition we can define the corresponding effective energy density,
pressure, heat flux and traceless stress-tensor, as measured by the observer
$u^{i}$, as
\begin{eqnarray}
&&\rho _{T}=\frac{1}{\left( u^{i}u_{i}\right) ^{2}}%
G_{ij}u^{i}u^{j}=-2c_{1}c_{3}+\frac{1}{r^{2}},  \label{TE.1} \\
&&p_{T}=\frac{1}{3}h^{ij}G_{ij}=2c_{1}c_{3}-\frac{1}{3r^{2}},  \label{TE.2}
\\
&&q^{i}=h^{ij}G_{jk}u^{k}=0,  \label{TE.3} \\
&&\pi _{\theta }^{\theta }=\pi _{\phi }^{\phi }=-\frac{1}{2}\pi _{r}^{r}=%
\frac{1}{3r^{3}},
\end{eqnarray}%
where
\begin{equation}
\pi _{ij}=(h_{i}^{r}h_{j}^{s}-\frac{1}{3}h_{ij}h^{rs})G_{rs},  \label{TE.4}
\end{equation}%
and $h_{ij}$ is the projective tensor
\begin{equation}
h_{ij}=g^{ij}-\frac{1}{\left( u^{i}u_{i}\right) }u^{i}u^{j}.
\end{equation}%
We mention that in the case of GR, namely when $f(T)=T$, such an effective
fluid does not exist. In this sense, $f(T)$ gravity resembles to $f(R)$
gravity, where the fact that  $f(R)\neq R$ gives rise to a 
curvature effective fluid \cite{Capozziello:2011et}. 

 From the above results,  we deduce that the obtained effective
energy 
momentum tensor is written as
\begin{equation}
T_{ij}=\hat{T}_{ij}+\tilde{T}_{ij}
\end{equation}%
where%
\begin{equation}
\hat{T}_{ij}=\hat{\rho}u_{i}u_{j}+\hat{p}h_{ij}
\end{equation}%
\begin{equation}
\tilde{T}_{ij}=\tilde{\rho}u_{i}u_{j}+\tilde{p}h_{ij}+\pi _{ij} \;.
\end{equation}%
Notice that we have made the corresponding splitting $\rho_{T}=\hat{\rho}+%
\tilde{\rho}$ and $p_{T}=\hat{p}+\tilde{p}$ with $\hat{p}=-\hat{\rho}%
=2c_{1}c_{3}$ and $\tilde{p}=-\frac{1}{3} \tilde{\rho}=-\frac{1}{3r^{2}}$.
Therefore, we conclude that the effective dark energy fluid,  due to the $f(T)$
terms, consists of two parts. The first, namely $\hat{T}_{ij}$, plays the
role of a cosmological constant, which is associated with the de
Sitter-Schwarzschild metric\footnote{%
The de Sitter-Schwarzschild solution is $A(r)=1-\frac{2\alpha}{r}-\frac{%
\Lambda }{3}r^{2}$ with $G^{i}_{j}=\mathrm{diag}\left( \Lambda ,\Lambda
,\Lambda ,\Lambda \right)$, $p=-\rho=-\Lambda$, $q_{i}=0$ and $\pi_{ij}=0$.}
with 
${\Lambda}=2c_{1}c_{3}$. 
The second part, namely $\tilde{T}_{ij}$
corresponds to a fluid with equation-of-state parameter equal to $-1/3$, and
is a pure effect of the $f(T)$ structure. 

In order to apply the above considerations for specific $f(T)$ forms, we
consider the following viable $f(T)$, motivated by cosmology \footnote{%
The $f(T)$ models of Refs.\cite{Ben09,Linn1} are consistent with the
cosmological data.}:

\begin{itemize}
\item Exponential $f(T)$ gravity \cite{Linn1}:
\begin{equation*}
f(T)=T+f_{0}e^{-f_{1}T},
\end{equation*}
where $f_{0}$ and $f_{1}$ are the two model parameters which are connected
via (\ref{yLL}), that is 
\begin{equation*}
f_{0}=\frac{4c_{1}c_{3}}{8f_{1}c_{1}c_{3}+1}\exp \left(
4f_{1}c_{1}c_{3}\right) .
\end{equation*}

\item A sum of two different power law $f(T)$ gravity:
\begin{equation*}
f(T)=T^{m}+f_{0}T^{n},
\end{equation*}
where from (\ref{yLL}) we have
\begin{equation*}
f_{0}=\frac{1-2m}{2n-1}\left( 4c_{1}c_{3}\right) ^{m-n } \;.
\end{equation*}%
Note that in the case of $m=1$ we recover the $f(T)$ model by Bengochea \&
Ferraro \cite{Ben09}.
\end{itemize}

Let us make a comment here. Recently, it has been showed \cite{BasNes13}
that the above viable $f(T)$ models may be written as perturbations around
the concordance $\Lambda$CDM model, which means that the corresponding
Hubble function of these $f(T)$ models can be given in terms of the $\Lambda$%
CDM Hubble parameter. Interestingly enough, in the current paper, we find
that the spherically symmetric solutions of the above $f(T)$ models smoothly
includes the Schwarzschild solution [see Eq.(\ref{ALLa})]. Generally, we
are interested in  viable $f(T)$ models, since these models can describe
the matter and dark energy eras, being consistent with the observational
data (including Solar System tests), and finally they have stable
perturbations. Although these necessary analyses have not yet been performed
for all the available $f(T)$ models, a failure of a particular model to pass
one of these tests is sufficient to exclude it. In this respect, we
plan to investigate in a forthcoming paper the performance of our spherical
solutions against the Solar System tests, aiming to impose constraints on
the free parameters.   

\subsection{ The case $f(T)=T^{n}$}

Based on considerations at  cosmological scales, it has been found by
Basilakos et al. \cite{BB1} that the $f(T)=T^{n}$ gravity models suffer for
two basic problems. The first is associated with the fact that the
deceleration parameter is constant, that is it never changes sign, and
therefore the universe always accelerates or always decelerates, depending
on the value of $n$. Secondly, the growth rate of cosmic structures remains
always equal to unity, implying that the recent growth data disfavor the $%
f(T)=T^{n}$ gravity. Despite the above caveats, for completeness in this
subsection we provide the analytical solutions for the spherically symmetric
geometry. In the $f(T)=T^{n}$ case, the field Eqs. (\ref{Lf.04}), (\ref%
{Lf.06}), (\ref{HLf.06}) and the torsion scalar (\ref{L4S01}) give rise to
the following dynamical system:
\begin{eqnarray}
&&T=4x^{\prime }y^{\prime},  \label{L4S1} \\
&& 4nT^{n-1}x^{\prime }y^{\prime }-\left( 1-n\right) T^{n}=0,  \label{L4S2}
\\
&& x^{\prime \prime }+\left( n-1\right) x^{\prime }T^{-1}T^{\prime }=0,
\label{L4S3} \\
&& y^{\prime \prime }+\left( n-1\right) y^{\prime }T^{-1}T^{\prime }=0\;.
\label{L4S4}
\end{eqnarray}
It is easy to show that combining Eq.(\ref{L4S1}) with the Hamiltonian (\ref%
{L4S2}), we can impose constraints on the value of $n$, namely $n=1/2$.
Under this condition, solving the system of Eqs. (\ref{L4S3}) and (\ref{L4S4}%
) we arrive at the solutions
\begin{eqnarray}
&&x(\tau)=\frac{\sigma(\tau)^{3}}{3}+c_{\sigma}, \\
&& y(\tau)=\frac{\sigma(\tau)^{3}}{3},  \label{L4S4b}
\end{eqnarray}
where $c_{\sigma}$ is the integration constant. Now using (\ref{L4S001}) we
derive $a$,$b$ as
\begin{eqnarray}  \label{L4S44}
&&b(\tau)=\sigma(\tau), \\
&& a(\tau)=\sqrt{\frac{2\left[\sigma^{3}(\tau)+3c_{\sigma}\right] }{%
3\sigma(\tau)}}.
\end{eqnarray}%
Using the coordinate transformation $\sigma(\tau)=r$, which implies $%
\tau=F(r)$ [with $F(\sigma(\tau))=\tau$], and using simultaneously (\ref%
{L4S44}), the spherical metric (\ref{SS}) can be written as
\begin{equation}  \label{MM44}
ds^{2}=-A(r)dt^{2}+B(r)dr^{2}+r^{2} \left( d\theta +\sin ^{2}\theta
d\phi^{2}\right)\;,
\end{equation}
where
\begin{equation}  \label{AA44}
A(r)=\frac{2}{3}r^{2}+\frac{2c_\sigma}{r} =\lambda_{A}(1-\frac{r_{\star}}{r}%
)R(r),
\end{equation}
with $\lambda_{A}=\left( \frac{8c^{2}_{\sigma}}{3}\right)^{1/3}$ 
and
\begin{equation}  \label{BB44}
B(r)=\frac{F_{,r}^{2}}{A(r)r^{4}}.
\end{equation}
The functional form of the distortion parameter $R(r)$ is given by relation (%
\ref{RLLa}), in which the characteristic distance becomes $%
r_{\star}=(-3c_{\sigma})^{1/3}=\left( \frac{3\lambda_{A}}{2}\right)^{1/2}$,
implying $c_{\sigma}<0$.

 Furthermore, considering the comoving observers  $\left(
u^{i}u_{i}\right) =-1$ , we can write the Einstein tensor components
as 
\begin{eqnarray*}
G_{t}^{t} &=&-\frac{r}{3F_{,r}^{3}}\left[ 4rF_{,rr}\left( r^{3}+3c_{\sigma
}\right) -2F_{,r}\left( 7r^{3}+12c_{\sigma }\right)\right.  \\
&+&\left.\frac{3}{r^{3}} F_{,r}^{3}\right]  \\
G_{r}^{r} &=&\frac{2r^{4}}{F_{,r}^{2}}-\frac{1}{r^{2}} \\
G_{\theta }^{\theta } &=&G_{\phi }^{\phi }=-\frac{1}{3}\frac{r}{F_{,r}^{3}}%
\left[ rF_{,rr}\left( 4r^{3}+3c_{\sigma }\right) -F_{,r}\left(
14r^{3}+6c_{\sigma }\right) \right] .
\end{eqnarray*}%

 Similarly, based on the first equalities of
(\ref{TE.1})-(\ref{TE.4}), we
provide the corresponding fluid components  
\begin{equation}
\rho _{T}=\frac{4r^{2}F_{,rr}}{F_{,r}^{3}}\left( \frac{1}{3}r^{3}+c_{\sigma
}\right) -\frac{2r}{F_{,r}^{2}}\left( \frac{7}{3}r^{3}-4c_{\sigma }\right) +%
\frac{1}{r^{2}}  \label{TEE.1}
\end{equation}%
\begin{equation}\label{TEE.2}
p_{T}=-\frac{2}{3}\frac{r^{2}F_{,rr}}{F_{,r}^{3}}\left( \frac{4}{3}%
r^{3}+c_{\sigma }\right) +\frac{2}{3}\frac{r}{F_{,r}^{2}}\left( \frac{17}{3}%
r^{3}+2c_{\sigma }\right) -\frac{1}{3r^{2}}
\end{equation}
\begin{equation*}
\pi _{,r}^{r}=\frac{2}{3}\frac{r^{3}F_{,rr}}{F_{,r}^{2}}\left( \frac{4}{3}%
r^{3}+c_{\sigma }\right) -\frac{2}{3}\frac{1}{F_{,r}^{2}}\left( \frac{8}{3}%
r^{3}+2c_{\sigma }\right) -\frac{2}{3r^{2}}
\end{equation*}%
\begin{equation*}
\pi _{\theta }^{\theta }=\pi _{\phi }^{\phi }=-\frac{1}{2}\pi _{r}^{r}\;
\end{equation*}%
\begin{equation*}
q^{i}=0
\end{equation*}

Finally, if we desire to construct an effective fluid that 
obeys a barotropic equation of state $p_{T}=(\gamma -1)\rho _{T}$ 
(frequently used in cosmological studies), then using Eqs.(\ref{TEE.1}), 
and (\ref{TEE.2}), we need to write $F_{,r}$ as 
\begin{equation}
F_{,r}=\frac{1}{\sqrt{Z\left( r\right) }}J\left( r\right) ^{\frac{%
3\gamma }{3\gamma -1}}r^{2}  \label{FEE2},
\end{equation}%
where
\begin{equation}
Z(r)=3\left( 3\gamma -2\right) \int J\left( r\right) ^{\frac{1}{3\gamma -1}%
}+F_{1}\;,
\end{equation}%
and
\begin{equation}
J(r)=(6\gamma -2)r^{3}+(18\gamma -15)c_{\sigma }.
\end{equation}%
From the above functions, it is clear that, in order to have a real
solution, the corresponding $\gamma$ parameter has to obey the restriction $%
\gamma >2/3$. To this end, inserting Eq.(\ref{FEE2}) into Eq.(\ref{BB44}) we
obtain
\begin{equation}
B(r)=\frac{J\left( r\right) ^{\frac{6\gamma }{3\gamma -1}}}{Z\left( r\right)
A(r)},
\end{equation}%
where the function $A(r)$ is given by (\ref{AA44}) and $F_{1}$ is the
constant of integration. 

  We mention here that such further terms could describe interesting
effects if coupled with an equation of state of the form
$p_m=K_0\rho_m^\gamma$ where $p_m$ and $\rho_m$ are the relative quantities
related to  standard matter fluids. As in the case of $f(R)$ gravity,
anomalous stars could be addressed by constructing modified Lan\'e - Emden
equations, where further geometric terms play a relevant role (see e.g.
\cite{star1}).
In particular, the above discussion could be useful in order to deal with 
anisotropic deformations of neutron star  instead of searching for exotic
form of matter \cite{anisotropic}. As discussed in \cite{neutron1,neutron2}
for  $f(R)$ gravity, geometric pressure terms, inserted in the
Tolman-Oppenheimer-Volkoff solution, could account for the larger effective
masses of some neutron stars, recently observed \cite{PSR1,PSR2},  that
escape the standard GR interpretation. 

\section{Conclusions}

\label{conclusions}

 In Basilakos et al. \cite{BB1} we have utilized the Noether symmetry
method in order to investigate the main properties of the $f(T)$ modified
gravity in the flat FLRW cosmology.  In this work, we present a complete
Noether symmetry analysis in the framework of $f(T)$ gravity. Specifically,
considering $f(T)$ gravity embedded in the static spherically symmetric
spacetime, we provide a full set of Noether symmetries for the related
minisuperspaces. Then we compute new analytical solutions for various $f(T)$
models. Interestingly, we find that the $f(T)$ static spherically symmetric
spacetime is written in terms of the well known Schwarzschild spacetime,
modified by a distortion function that depends on a characteristic radius.
We mention that the obtained solution classes are more general and cannot be
obtained by the usual solutions methods. Obviously, the combination
of the work by  Basilakos et al. \cite{BB1} with the current article provide
a complete investigation of the Noether symmetry approach in $f(T)$ gravity
at FLRW and spherical levels respectively. 

From a genuine physical point of view, this means that $f(T)$ gravity could
be a reliable approach in order to deal with several open issues in
astrophysics and cosmology. In a forthcoming paper, we will consider in
details such applications.

\begin{acknowledgments}
 SB acknowledges the hospitality of the Astrophysics
Group of Dipartimento di Fisica, Universit\'a di Napoli "Federico
II'', where a major part of this work was done. SB acknowledges support by
the Research Center for Astronomy of the Academy of Athens in the context of
the program {\it ``Tracing the Cosmic Acceleration''}.  The research of ENS
is implemented within the framework of the Action «Supporting Postdoctoral
Researchers» of the Operational Program ``Education and Lifelong Learning''
(Action's Beneficiary: General Secretariat for Research and Technology), and
is co-financed by the European Social Fund (ESF) and the Greek State.
SC acknowledges the support by INFN (iniziative specifiche TEONGRAV and QGSKY).
 KA
acknowledges support by the Research Institute for Astronomy and Astrophysics
of Maragha (RIAAM) under research project No. 1/2782-59.
\end{acknowledgments}

\end{document}